\documentclass{article}
\usepackage{graphicx} 
\usepackage{booktabs}
\usepackage{amssymb}
\usepackage{amsmath}
\usepackage{amsthm}

\usepackage{xspace}
\newtheorem{theorem}{Theorem}
\newtheorem{definition}{Definition}

\usepackage{blindtext}
\usepackage{geometry}
\usepackage{caption}
\usepackage{subcaption}
\usepackage{float}
\usepackage[numbers]{natbib}
\usepackage{url}

\newcommand{\BibTeX}{\rm B\kern-.05em{\sc i\kern-.025em b}\kern-.08em\TeX}

\title{Identifying the Source of Information Spread in Networks via Markov Chains}

\author{Yael Sabato\\The Open University of Israel\\yael.sabato@openu.ac.il \and Amos Azaria\\Ariel University\\amos.azaria@ariel.ac.il \and Noam Hazon\\Ariel University\\noamh@ariel.ac.il }

\begin{document}
\date{}
\maketitle

\begin{abstract}
Nowadays, the diffusion of information through social networks is a powerful phenomenon. One common way to model diffusions in social networks is the Independent Cascade (IC) model. Given a set of infected nodes according to the IC model, a natural problem is the source detection problem, in which the goal is to identify the unique node that has started the diffusion. Maximum Likelihood Estimation (MLE) is a common approach for tackling the source detection problem, but it is computationally hard.

In this work, we propose an efficient method for the source detection problem under the MLE approach, which is based on computing the stationary distribution of a Markov chain. Using simulations, we demonstrate the effectiveness of our method compared to other state-of-the-art methods from the literature, both on random and real-world networks.
\end{abstract}


\section{Introduction}

In the age of social media, the spread of information and infection through networks is a significant phenomenon. Understanding the dynamics of information spread and identifying its origin are important for a wide range of applications, including marketing, public health, and identification of fake news. The Independent Cascade (IC) is a common model of the spread of information in a social network \cite{goldenberg2001talk}. In the IC model, the process of diffusion concerns a message that is propagated through the network. Every connection between two friends is associated with a probability; this value determines the probability that if the first user shares the message, the second user will share the message with her friends as well. As commonly occurs in the spread of fake news, the diffusion process starts with a single initial source. A natural goal is that given a set of users who shared a specific message, to seek the unique source that started the diffusion.


There are many approaches to finding the source of a diffusion in the literature, each assuming different spreading models and various amounts of knowledge of the network parameters (for example, \cite{lappas2010finding,gomez2012inferring,zhang2017markov,kumar2017temporally}). When the probabilities associated with the connections are known or can easily be estimated, a natural mathematical approach for finding the source of the diffusion is the Maximum Likelihood Estimation (MLE) principle. According to the MLE principle, one should compute the likelihood of each user being a source, and output the user with the maximum likelihood.

The first to formalize the computational problem of finding the source of a diffusion in a network in the IC model are \citeauthor{lappas2010finding} \cite{lappas2010finding}. 
They show that for arbitrary graphs, the source detection problem is not only NP-hard to find but also NP-hard to approximate. Therefore, they propose an efficient heuristic, but it does not utilize the MLE principle. \citeauthor{zhai2015cascade} \cite{zhai2015cascade} present a heuristic that utilizes the MLE principle, and they further show that their heuristic outperforms the heuristic of \cite{lappas2010finding}. However, their heuristic requires extensive computation. In addition, they note that ``although the IC model is popular in social network research, finding source in the IC model is rarely studied''. Recently, \citet{amoruso2020contrasting} provide a strong heuristic for finding the source of a diffusion in a network in the IC model.

In this paper, we propose an efficient method that uses the MLE principle for source detection in the IC model.
Our method is based on computing the stationary distribution of a Markov chain and is inspired by \cite{kumar2017temporally}. Specifically, we recognize that if we represent the social network as a weighted directed graph,
the diffusion in the IC model induces a tree that spans the set of users who shared the message, and the root of the tree is the user who initiated the diffusion. In addition, the tree is associated with a weight, which is equal to the product of the weights of its edges.
In order to estimate the probability of a specific user being the source, we would like to sum the weights of all spanning trees rooted at this user. However, directly considering all trees is computationally expensive. Therefore, we propose converting the social network into a Markov chain, and the Markov chain tree theorem \cite{leighton1986estimating}, 
allows us to compute the sum of the weights of all spanning trees rooted at each user in polynomial time. 
We consider two approaches for converting the social network to a Markov chain---the \emph{self-loops} and the \emph{no-loops} methods. The self-loops method is (arguably) more intuitive and easier to comprehend, but both methods result in the exact same solution and take a similar time to arrive at this solution when using a direct calculation of the stationary distribution.

For evaluating the effectiveness of our approach, we use $14$ types of random
graphs, and sample $1000$ graphs from each type. In addition, we evaluate the effectiveness of our approach on $8$ real-world networks, including a portion of Digg, Facebook, and Twitter.
We first show that the no-loops method outperforms the self-loops method when using a random walk to estimate the stationary distribution. Since both methods result in the exact same solution when using a direct calculation of the stationary distribution, we conclude that the no-loops method is superior to the self-loops method.
We then compare the performance of the no-loops method with several baseline methods, including the method proposed by \cite{zhai2015cascade} and \cite{amoruso2020contrasting}. 
Our experiments indicate that the no-loops method outperforms all of the other methods, both on random graphs and real-world networks.

\section{Related Work}


Viral spread of information through social networks inspired various researchers. The most studied problem is the Influence Maximization (IM) problem, in which the goal is to find the most influential node (or set of nodes) in a social network \cite{kempe2003maximizing}. The source detection problem, which we focus on, is (in a way) the inverse of the IM problem: instead of finding the node that will result in the maximal diffusion (i.e., the IM problem), we are given the outcome of a diffusion that has occurred, and we would like to find the source node that initiated it.

There have been multiple approaches for attempting to solve the source detection problem, with different models and with different settings, as can be seen in the following reviews \cite{jiang2016identifying,shelke2019source,jin2021schemes}.
Indeed, most of the prior work consider epidemic models such as the susceptible–infected (SI) model, and the susceptible–infected–recovery (SIR) model. For example, \citet{shah2010detecting} and \citet{shah2011rumors} consider the SIR and SI models, respectively, and present algorithms that are based on rumor centrality. 
\citet{agaskar2013fast} also consider the SI model, and propose an algorithm that is based on geodesic distances on a randomly-weighted version of the network.
\citet{kumar2017temporally} consider the SI model, but with an assumption that some information regarding the edges that participate in the diffusion is provided. 
Since they consider the SI model, they assume that the network is undirected and the edges are unweighted. That is, each user is equally likely to get infected (i.e., receive a message) by each of her neighbors.

The IC model is the classic information-propagation model, and is widely used in social network research. However, only a few papers study the source detection problem with the IC model.
\citet{lappas2010finding} are the first to formulate the computational problem of finding the source of a diffusion in the IC model. They propose an efficient heuristic, which is based on dynamic programming, but they do not utilize the MLE principle.
\citet{zhai2015cascade} also consider the IC model, and show that finding the source of a diffusion is $\#P$-complete. Therefore, they develop a heuristic that utilizes the MLE principle, and is based on a Markov chain. However, their use of the Markov chain is completely different from ours. Specifically, they define the states of the Markov chain as different samples of the network. Their algorithm then performs random walks on the Markov chain, and for each sample it computes the set of all possible source nodes. Although their algorithm outperforms the heuristic of \citet{lappas2010finding}, it requires extensive computation to do so, and it does not perform well when compared to our method.   
Similar to the work of \citet{zhai2015cascade}, \citet{zhang2017markov} use a Markov chain in which the states are different samples of the network. However, their model of diffusion is the linear threshold model, instead of IC. 
\citet{tong2016effector} consider the IC model and utilize the MLE approach, but they study a slightly different problem, the effector detection problem. That is, instead of trying to find the source of a diffusion, their goal is to find the node that can best explain the current state of the other nodes. Therefore, given the same input, the solution to the effector detection problem may be different from the solution to the source detection problem.
\citet{berenbrink2023inference} consider the IC model and assume a fixed activation probability for all edges. They provide strong information-theoretic results for acyclic (undirected) graphs. They also suggest a simple heuristic and demonstrate its performance on several random graphs.

Several works have considered the problem of finding multiple sources of diffusion \cite{zhu2017catch,wang2017multiple,amoruso2020contrasting}. However, we believe that, in practice, there usually is only a single source. For example, in epidemics, there is a single patient zero, and it is very unlikely that two people (or more) will simultaneously develop the exact same disease. Similarly, in fake news, there is a single author of the news, and in rumor spreading, typically one individual or entity initiates the rumor.

\section {Preliminaries}
\subsection{Directed Rooted Trees}
A directed rooted tree is a directed acyclic graph (DAG) whose underlying undirected graph is a tree, and one of its vertices has been designated the root.
An out-tree is a directed rooted tree, in which all the edges point away from the root. Similarly, an in-tree is a directed rooted tree, in which all the edges point toward the root. Clearly, we can convert an out-tree into an in-tree by reversing the directions of the edges. A spanning in-tree/out-tree is an in-tree/out-tree that spans all vertices of the underlying graph. 
\subsection{The Diffusion Model}
\label{sec:IC_model}
The research on the diffusion of information on social networks has considered several models. In this paper we focus on the independent cascade model (IC), which is the following.
There is a social network that is represented by a weighted directed graph $G_N=(V_N,E_N)$ with no self loops, where each user of the social network is represented by a node, every connection between two users is an edge, and the weights represent the influence probabilities. 
The process of diffusion concerns a message that is propagated through the social network. During this process, each node can either be inactive or become active.
The diffusion process starts with an initial source   
$v\in V_N$, which is the first active node.
The process then unfolds in discrete steps according to the following rule. 
Every node $v_i\in V_N$ that becomes active in step $t-1$, attempts to activate each currently inactive neighbor $v_j$ in step $t$, and only in step $t$. The probability that $v_i$ succeeds in activating an inactive neighbor $v_j$ is $p^{ij}$, which is the weight of the edge $(v_i,v_j)\in E_N$. We denote by $w_{in}(v_i)$ the weighted in-degree of a node $v_i$, $w_{in}(v_i) = \sum_j{p^{ji}}$.
If multiple neighbors of a vertex $v$ try to activate it at the same time, their attempts are considered in an arbitrary order. 
The process runs until no more activations occur. 

Note that since each active node is activated by a single parent, the active nodes and activating edges 
form an out-tree, which spans the set of active nodes, and the source node is the root of the tree.


\subsection{Markov Chain}
\label{sec:maarkov_chain}

A (discrete-time) Markov chain is 
an infinite sequence of discrete random variables $(X_i)_{i=0}^{\infty}$.
All variables have the same finite set of possible values, $S = \{s_1,s_2,\ldots,s_n \}$, which is called the state set of the Markov chain.
The variables of the sequence $(X_i)_{i=0}^{\infty}$ have the Markov property of \textit{forgetfulness}, that is, each variable $X_t$ is dependent only on the previous variable $X_{t-1}$, and is independent of all other previous variables. Moreover, all the variables have the same probability distribution.
Namely, for every $t>0$, $P(X_{t}=s_{i_t}|X_1=s_{i_1},X_2=s_{i_2},...,X_{t-1}=s_{i_{t-1}}) =P(X_{t}=s_{i_t}|X_{t-1}=s_{i_{t-1}})=P(X_{t+1}=s_{i_t}|X_{t}=s_{i_{t-1}})$, where $s_{i_j} \in S$. For a pair of states $s_i$ and $s_j$, let $q^{ij}=P(X_{t}=s_j|X_{t-1}=s_i)$. We call $q^{ij}$ the \textit{transition probability}. Note that $\sum_{j=1}^n{q^{ij}}=1$. 



A Markov chain can be represented as a weighted directed graph $G_M=(S_M,E_M)$, where each state is represented by a node. For clarity reasons, we refer to the nodes of $S_M$ as states.
There is an edge $(s_i,s_j) \in E_M$ if the transition probability $q^{ij}>0$, with a weight of $q^{ij}$.
A random walk on the graph $G_M$ is called a Markov process.

\subsubsection{Irreducibility}
A Markov chain is \textit{irreducible} if for each pair of states $s_i,s_j$, there are two directed paths $s_i \leadsto
s_j$ and  $s_j \leadsto s_i$ in $G_M$. That is, a Markov chain is irreducible if its graph, $G_M$, is strongly connected.

\subsubsection{Stationary Distribution}

If the Markov chain is irreducible
then the long run average number of visits
of the Markov process to any state $s_i$, converges to a number $\Pi_i$, 
regardless of the initial state. That is, for all 
$s_i\in S$ it holds that 
$\Pi_i = \lim_{k\rightarrow \infty}\frac{1}{k}\sum_{t=0}^{k-1} \mathbb{I}(X_t=s_i)$, 
where $\mathbb{I}(\cdot)$ is an indicator function. Moreover, $\Pi = (\Pi_1,$ $\Pi_2,$ $\ldots,$ $\Pi_n)$ is a probability distribution over the set of states $S$, which is called the \textit{Stationary Distribution}\footnote{Some works provide a slightly different definition for the stationary distribution, which requires that the Markov chain will also be ergodic \cite{batabyal2006markov}.}.
The stationary distribution can be computed in polynomial time \cite{levin2017markov}.
However, if $G_M$ is very large the exact computation of the stationary distribution may become impractical; in this case, it is common to estimate the stationary distribution by sampling, i.e., using random walks on the graph.


\subsection{Vector Notations}

\sloppy
Let $V$ be a vector, $V= (V_1, V_2, \ldots ,V_n)$. We denote by $\hat{V}$ the normalized vector, $\hat{V} = (\frac{V_1}{\sum_{i=1}^n V_i}, \frac{V_2}{\sum_{i=1}^n V_i}, \ldots ,\frac{V_n}{\sum_{i=1}^n V_i})$.


\section{Problem Statement} \label{sec: problem statement} 
In this section, we define the source detection problem and present the MLE principle, which provides a framework for addressing it.
\begin{definition}[Source Detection]
	Given a social network, $G_N=(V_N,E_N)$, and a set of active nodes $A\subseteq V_N$ at the end of a propagation process that is compatible to the IC diffusion model, find the source node. 
\end{definition}

We first note that a source node $v$ must be in $A$. Moreover, there must be a directed path from $v$ to each node in $A$. 
Let $A' \subseteq A$ be the set of all nodes that have directed paths to all nodes in $A$. Clearly, $A'$ is strongly connected.
In addition, if $A'$ is the singleton $\{v\}$ then $v$ is the source node. 

Our approach for identifying the source node is to follow the maximum likelihood principle \cite{myung2003tutorial}. That is, each node is associated with the probability that it is the source, and we select a node with the maximal probability. Formally, let $R_i$ be the event that $v_i$ is the source node, and let $\mathcal{A}$ be the event that the set $A$ is the set of active nodes; $\mathcal{A'}$ is defined similarly for $A'$. We would like to solve the following problem:

\begin{definition}[ML-Source]
	Given a social network, $G_N=(V_N,E_N)$, and a set of active nodes $A\subseteq V_N$ at the end of a propagation process according to the IC diffusion model, find the most likely source node $v^*$, i.e., $$
	v^* = \arg\max_{v_i \in V_N} P(R_i | \mathcal{A}).
	$$
\end{definition}
\noindent Note that $P(R_i | \mathcal{A}) = \frac{P(R_i,\mathcal{A})}{P(\mathcal{A})}$, and thus 
$$
\arg\max_{v_i \in V_N} P(R_i | \mathcal{A}) = \arg\max_{v_i \in V_N} P(R_i,\mathcal{A}).$$
%

Unfortunately, the ML-Source problem was shown to be computationally hard~\cite{zhai2015cascade}.
Indeed, the following brute-force procedure computes the exact value of $P(R_i,\mathcal{A})$, in exponential time.
Let $G_N[A] = (A, E(A))$ be the subgraph of $G_N$ induced by the set $A$. That is, $E(A)$ is the set of edges of $G_N$ that have both nodes in $A$.
We consider every subset $X\subseteq E(A)$, and each such $X$ is associated with a probability for its occurrence: 
$$
p(X)=\prod_{e\in X}p^e\cdot \prod_{e\in E(A)\setminus X} (1-p^e).
$$
Let $G_X$ be a graph, $G_X =(A,X)$. If there exists an out-tree in $G_X$ that spans $A$ and with the node $v_i$ as the root, then the probability $p(X)$ should be added to $P(R_i,\mathcal{A})$. Namely:
$$
P(R_i,\mathcal{A}) = \sum_{X\subseteq E(A)} I (X,v_i)\cdot p(X)
$$
where $I(X,v_i)$ is an indicator function that returns $1$ if the graph $G_X$ has a spanning out-tree with $v_i$ as the root, and $0$ otherwise.
In order to return the most likely source node, one should compute the above expression for every $v_i\in A$. Moreover, each $p(X)$ can be used more than once, in the case where $G_X$ has multiple spanning out-trees with several roots.

\section{The Markov Chain Approach}
\citeauthor{kumar2017temporally}~\cite{kumar2017temporally} suggested the Markov chain approach for estimating the probability of a node to be the source, in their setting (i.e., the SI model on an undirected graph, with an assumption that some information regarding the edges that participate in the diffusion is provided).
We adapt the Markov chain approach to our setting, as follows.

For a diffusion that started with a source node $v_i$, and
resulted in a set $A$ of active nodes,  
let $T_{i,A}$ be the corresponding spanning out-tree, and let $\mathcal{T}_{i,A}$ be the associated event. We denote by $w(T)$ the weight of a directed rooted tree, $w(T) = \prod_{e\in T}p^e$.
A good estimation of the probability of $\mathcal{T}_{i,A}$ is:
$$
P(\mathcal{T}_{i,A}) \approx P(R_i)\cdot w(T_{i,A}).
$$
Note that this is an estimation, since we ignore the edges that are not part of the spanning out-tree. 
In order to calculate the probability of a single node $v_i$ to be the source, we go over every spanning out-tree rooted at $v_i$ and sum the weights of those out-trees:

$$
P(R_i,\mathcal{A}) \approx \sum _{T_{i,A}\in OT_{i,A}} P(\mathcal{T}_{i,A}).
$$
where $OT_{i,A}$ is a set of all the out-trees of $G_N[A]$ that are rooted at $v_i$ and span $A$. 
Note that this summation is also an estimation, since the events $\mathcal{T}_{i,A}$ for every spanning out-tree are not independent (which can be fixed with an inclusion-exclusion calculation).


For  $v_i \in A'$, let $\Gamma_i =  \sum _{T_{i,A'}\in OT_{i,A'}} w(T_{i,A'})$, let $\Gamma =(\Gamma_1,$ $ \Gamma_2,$ $\ldots,$ 
$\Gamma_{|A'|})$, 
We assume that the \textit{prior} probability, $P(R_i)$, is equal for every $v_i\in V_N$. It is also enough to consider $A'$ instead of $A$ since $P(R_i,\mathcal{A}) \propto P(R_i,\mathcal{A'})$ (according to \cite{kumar2017temporally}). 
We get that
\begin{equation}
	\label{eq: v* by spreading patterns}    
	v^* \approx \arg\max_{v_i \in A'} \Gamma_i.
\end{equation}

Based on this formulation, a naive approach is to compute for each $v_i\in A'$ the set $OT_{i,A'}$, (using an algorithm for finding all spanning out-trees, e.g. \cite{gabow1978finding}), and to return the node $v_i$ that maximizes $\Gamma_i$. We refer to this approach as the out-tree counting method.
Clearly, the out-tree counting method is (also) computationally expensive, as the size of $OT_{i,A'}$ is most likely exponential in $|A'|$. 
We thus propose to use the following theorem 
\cite{leighton1986estimating}:

Given a finite state irreducible Markov chain as a directed graph $G_M=(S_M,E_M)$, 
Let $w(T) = \prod_{e\in T} p^e$ be the weight of a spanning in-tree $T \subseteq E_M$. Let $IT_{i,S_M}$ be the set of all in-trees in $E_M$ that have $s_i$ as their root and span $S_M$. Let $\Psi_i:= \sum _{T\in IT_{i,S_M}}w(T)$, let $\Psi = (\Psi_1, \Psi_2, \ldots, \Psi_n)$. 

\begin{theorem} \label{TreeThm}
	(Markov chain tree theorem)
	Given a finite state irreducible Markov chain, For every $s_i\in S_M$, the unique Stationary Distribution $\Pi_i$ is equal to $\hat{\Psi}_i$. namely:
	\begin{equation}
		\forall s_i\in S_M,  \Pi_i = \hat{\Psi}_i = \frac{\Psi_i}{\sum _{j=1}^n\Psi_j}    
	\end{equation}
\end{theorem}

Our approach is based on exploiting the structural similarity between $\Gamma$ and $\Psi$. Indeed, let $G_N[A']$ be the graph $G_N$ induced on the set $A'$, then for every $1\leq i\leq |A'|$, $\Gamma_i$ is the summation of weights of spanning out-trees in $G_N[A']$,
and $\Psi_i$ is the summation of weights of spanning in-trees in a Markov chain.
We thus first convert the social network $G_N[A']$ 
into a Markov chain $G_M$. 
This conversion includes the inversion of all the edges. That is, each node $v_i\in G_M$ is represented by a state $s_i$,
and each edge $(v_i,v_j)$ in $G_N[A']$ is converted to an edge $(s_j,s_i)$ in $G_M$. 
Therefore, every spanning out-tree in $G_N[A']$ corresponds to a spanning in-tree in $G_M$. We then compute the complete stationary distribution $\Pi$, of the Markov chain $G_M$, obtaining $\hat{\Psi}$
(by Theorem \ref{TreeThm}).
We then use $\hat{\Psi}$ to restore $\hat{\Gamma}$. Finally, we output the node with the maximal $\hat{\Gamma}$ value. 

Observe that converting the social graph into a Markov chain must be performed carefully. Specifically, it
requires that the transition probabilities are valid, i.e., for each state $s_i$, $\sum_{j=1}^n q^{ij} = 1$.
Moreover, it requires that the $\hat{\Gamma}$ values can be efficiently restored from $\hat{\Psi}$.

The stationary distribution $\Pi$ can be computed in polynomial time, and so can finding the set $A'$ (see for example \cite{sharir1981strong}), and the conversion of the social graph into a Markov chain. Therefore, our heuristic can be efficiently computed.

%

\section{Conversion Methods}  \label{sec: conversion method}

%
%

Now we detail how to produce the Markov chain graph $G_M=(S_M,E_M)$.
Each node $v_i\in A'$ is represented by a state $s_i\in S_M$, and each directed edge $(v_i,v_j)\in E(A')$  is represented by the reversed edge $(s_j,s_i)\in E_M$. We get that in the Markov chain, each edge is pointing from a node to all its possible activators. Therefore, a naive approach for converting the social graph into a Markov chain is to divide the weights of each edge of the Markov chain, by the sum of all weights of the incoming edges of the original node. That is, for each node $v_i\in A'$ and each edge $(s_j,s_i) \in E_M$, set $q^{ji} = \frac{p^{ij}}{w_{in}(v_j)}$. 
%

\begin{figure}
	\centering
	\begin{subfigure}[b]{0.40\columnwidth}
		\centering
		\includegraphics[width=\linewidth]{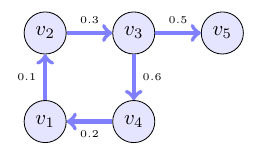}
		\caption{}
		\label{fig:simple_ex}
	\end{subfigure}
	\quad 
	\begin{subfigure}[b]{0.40\columnwidth}
		\centering
		\includegraphics[width=\linewidth]{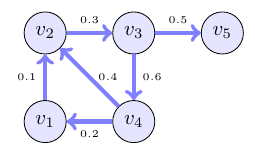}
		\caption{}
		\label{fig:complex_ex}
	\end{subfigure}
	\caption{Graph examples.}
	\label{fig:two graphs}
\end{figure}

However, 
merely normalizing the edge probabilities is 
not enough, since we lose the distinction between nodes with different $w_{in}(\cdot)$ values. (And clearly, \textit{ceteris paribus}, a node with a low $w_{in}(\cdot)$ value is more likely to be the source). 
For example, consider the social network in Figure~\ref{fig:simple_ex}.
In this example, the possible sources are $A'=\{v_1,v_2,v_3,v_4\}$, and the naive normalization assigns a weight of $1$ to all the edges of the Markov chain. This results in a stationary distribution in which $\Pi_1 = \Pi_2 = \Pi_3 = \Pi_4 = 0.25$~\footnote{The Markov chain is $S_M=\{s_1,s_2,s_3,s_4\}$ , $E_M=\{(s_1,s_4),$ $(s_4,s_3),$ $(s_3,s_2),$ $(s_2,s_1)\}$ and all the transition probabilities equal $1$, therefore the stationary distribution is $\Pi=(0.25,0.25,0.25,0.25)$.}. 
However, when computing the exact probabilities, we get that 
$p(R_1|\mathcal{A'})=p(\mathcal{A'}|R_1) \cdot p(R_1)/p(\mathcal{A'}) = 0.1\cdot 0.3\cdot 0.6 \cdot \frac{1}{z}$, where $z$ is the same for every $R_i$. Similarly, $p(R_2|\mathcal{A'})=0.3\cdot 0.6\cdot 0.2 \cdot \frac{1}{z}$,
$p(R_3|\mathcal{A'})=0.6\cdot 0.2\cdot 0.1 \cdot \frac{1}{z}$,
$p(R_4|\mathcal{A'})=0.2\cdot 0.1\cdot 0.3 \cdot \frac{1}{z}$. Thus, $z= 0.072$ and the vector of probabilities is $(0.25,0.5,0.167,0.083)$, in which  $p(R_i|\mathcal{A'})$ is in the $i$-th position. That is, $v_2$ is much more likely than any other node to be the source, but the naive approach fails to identify $v_2$ as the most likely source node. 

We thus present two methods for converting the social graph into a Markov chain, which consider (among other things) the difference between the $w_{in}(\cdot)$ values.

\subsection{The Self-Loops Method}


%


In our first approach, \textit{Self-loops}, after converting all the edges of $G_N[A']$, we add self-loops to all states. This allows us to normalize the edge probabilities by dividing all of the weights by the same number. Specifically, let $max_{in} = \max_{v_i\in A'} w_{in}(v_i)$ (We compute $w_in$ in the graph $G_N[A']$).
The self-loops method works as follows;
\begin{enumerate}
	\item Convert each node $v_i \in G_N[A']$ into a state $s_i$ and each edge $(v_i,v_j)$ into a reversed edge $(s_j,s_i)$ with $q^{ji} = \frac{p^{ij}}{max_{in}}$.
	\item For each $s_i$, add a self loop $(s_i,s_i)$ with $q^{ii}=\frac{max_{in}-w_{in}(v_i)}{max_{in}}$
	\item Compute the stationary distribution $\Pi$.
	\item $\hat{\Gamma}$ is assigned the values of $\Pi$. 
\end{enumerate}




Clearly, the transition probabilities are valid. We need to show that $\hat{\Gamma}$ is restored correctly. 

\begin{figure}
	\begin{subfigure}[b]{0.45\columnwidth}
		\includegraphics[width=\linewidth]{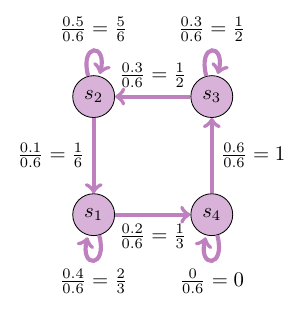}
		\caption{}
		\label{fig simple self loops markov chain}
	\end{subfigure}
	\quad
	\begin{subfigure}[b]{0.45\columnwidth}
		\includegraphics[width=\linewidth]{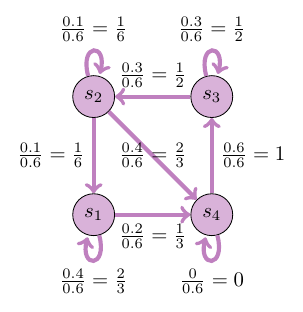}
		\caption{}
		\label{fig self loop mc}
	\end{subfigure}
	\caption{The Markov chains that are obtained by applying the self-loops method on the graphs of Figure \ref{fig:two graphs}.}
\end{figure}

	
	

\begin{theorem}
	\label{SelfLoop}
	(Self-loops method)
	
	For the self-loops method, for every $1\leq i \leq |A'|$ it holds that $\Psi_i =\alpha \cdot \Gamma_i$,
	where $\alpha$ is a constant.
	In other words, For every node $v_i\in G_N[A']$, the sum of weights of all out-trees spanning $A'$ and rooted at $v_i$,
	is proportional to the sum of weight of all in-trees in $G_M$ that span all states and are rooted at the corresponding state $s_i$. 
\end{theorem}

\begin{proof}
	Observe that every spanning out-tree $T \subseteq G_N[A']$, with weight $w(T)=\prod_{e\in T}p^e$ has a corresponding spanning in-tree $T' \subseteq G_M$ with weight $w(T')$. In addition,
	in-trees do not contain self loops, and have exactly $n-1$ edges (where $n=|A'|$). Therefore,
	\[
	w(T')=\prod_{e\in T'}q^e = 
	\prod_{e\in T}\frac{p^e}{max_{in}} 
	\]
	\[
	=\frac{1}{(max_{in})^{n-1}}\cdot \prod_{e\in T}p^e 
	=\alpha \cdot w(T).
	\]
	Thus,
	\[
	\Psi_i = \sum_{T'\in IT_{i,S_M}}w(T') = \alpha \cdot \sum_{T\in OT_{i,A'}} w(T)=\alpha \cdot \Gamma_i.
	\]
\end{proof}

Clearly, it can be concluded that $\hat{\Psi} = \hat{\Gamma}$.
Note that the stationary distribution $\Pi$ that is calculated by the self-loops method is equal to $\hat{\Psi}$ (by Theorem \ref{TreeThm}), and is also equal to $\hat{\Gamma}$. Therefore, by using the self-loops method for the conversion and selecting the node $v_i$ with the maximal $\hat{\Gamma}_i$, we obtain a good estimation for $v^*$. 



To demonstrate the self-loops method and Theorem \ref{SelfLoop}, we consider the social network given in Figure \ref{fig:complex_ex}, which is slightly more complex than the graph in Figure \ref{fig:simple_ex}, as it includes an additional edge between $v_4$ and $v_2$.
Now, assume that the set of active nodes is $A= \{v_1,v_2,v_3,v_4,v_5\}$.
Clearly, $A'=\{v_1,v_2,v_3,v_4\}$, 
The self-loop method computes
the Markov chain that is shown in Figure \ref{fig self loop mc},
and the stationary distribution of this Markov chain is $\Pi = (0.125,$ $0.25,$ $0.417,$ $0.208)$. Recall that $\Pi= \hat{\Psi}$, (according to Theorem~\ref{TreeThm}), and $\hat{\Psi}=\hat{\Gamma}$ (according to Theorem~\ref{SelfLoop}).
Indeed, using the out-tree counting method for directly calculating the values of $\Gamma$ leads to the same result. Specifically, $v_1$ has one possible spanning out-tree with $\Gamma_1 = 0.1\cdot 0.3 \cdot 0.6 = 0.018$. $v_2$ has one possible spanning out-tree with $\Gamma_2= 0.3\cdot 0.6\cdot 0.2=0.036$. $v_3$ has two possible spanning out-trees with $\Gamma_3 =0.6\cdot 0.2 \cdot 0.1 +0.6\cdot 0.2 \cdot 0.4=0.06$ 
and $v_4$ has two possible spanning out-trees with $\Gamma_4 =0.2\cdot 0.1 \cdot 0.3 +0.2\cdot 0.4 \cdot 0.3=0.03$ 
Therefore, $\Gamma = (0.018,$ $0.036,$ $0.06,$ $0.03)$, and $\hat{\Gamma}$ equals $\Pi$. Overall, the self-loops method outputs the vertex $v_3$ as the most likely source node.





Note that the \textit{precise} brute force calculation outputs the values $(0.1315,0.2631,$ $0.4035,$ $0.2017)$, and thus it also determines that $v_3$ is the most likely source node. Furthermore, the correlation between the exact probabilities and $\Pi$ is $0.9966$. 


Returning to the example in Figure \ref{fig:simple_ex}, the corresponding Markov chain that is obtained by the self-loops method is shown in figure \ref{fig simple self loops markov chain}, and the stationary distribution for this Markov chain is $\Pi=(0.25,$ $ 0.5,$ $ 0.167,$ $0.083)$. 
That is, in this example the self-loops method finds the exact probabilities, since for every sub-graph of $G_N[A']$ there is at most one spanning out-tree. 




\subsection{The no-loops Method} \label{sec:no-loop}
Recall that the self-loops method returns the exact values of $\hat{\Gamma}$ when $\Pi$ is computed directly. However, when $\Pi$ is estimated by sampling, the addition of the self loops to the graph might require longer random walks, as many of the random steps are ``wasted'' on the self loops. This, in turn, may affect the accuracy of the estimation of $\Pi$.
Therefore, we present our second method, \textit{no-loops}, which does not add self-loops to the states. Instead, it first converts the graph and computes the edge probabilities using the naive method. Once the stationary distribution $\Pi$ is computed, the no-loops method restores the correct $\hat{\Gamma}$ values from $\Pi$ by dividing each $\Pi_i$ by $w_{in}(v_i)$ and normalizing the result.

Specifically, the no-loops method is executed as follows:
\begin{enumerate}
\item Convert each node $v_i \in G_N[A']$ into a state $s_i$ and each edge $(v_i,v_j)$ into a reversed edge $(s_j,s_i)$ with $q^{ji} = \frac{p^{ij}}{w_{in}(v_j)}$. 
	\item Compute the corresponding stationary distribution $\Pi$.
	\item Let $\Pi^{corr} =$ $ (\Pi^{corr}_1,$ $ \ldots, \Pi^{corr}_{|A'|})$, where for every $1\leq i\leq |A'|$, $\Pi^{corr}_i =$ $\frac{\Pi_i}{w_{in}(v_i)}$.
	\item $\hat{\Gamma}$ is assigned the values of $\hat{\Pi}^{corr}$.
	
\end{enumerate}

We now show that $\hat{\Gamma}$ is restored correctly.  Indeed, the no-loops method converts the social network to a Markov chain using the naive method. We show that with this conversion, the weight of each spanning out-tree is divided by a value that depends only on the root node.
Therefore, in order to restore $\hat{\Gamma}$ we must multiply each $\Pi_i$ by this value. Indeed, as we show, instead of multiplying by this value, it is sufficient to divide by $w_{in}(v_i)$.

\begin{theorem}\label{NoLoop}
	(No-loops method)
	
	For the no-loops method, for every $1\leq i \leq |A'|$ it holds that $\Pi_i^{corr} 
	=\alpha \cdot \Gamma_i$,
	where $\alpha$ is a constant.
	
	
\end{theorem}

\begin{proof}
	Let $T$ be a spanning out-tree in $G_N[A']$, rooted at $v_r$ with weight $w(T) =\prod _{(v_i,v_j)\in T}p^{ij}$.
	Additionally, let $T'$ be the spanning in-tree in $G_M$ that corresponds to $T$.
	Each edge $(s_j,s_i) \in T'$ has a weight $q^{ji} = \frac{p^{ij}}{w_{in}(v_j)}$. 
	Therefore, the weight of $T'$ is:
	
	\[
	w(T') = \prod _{(s_j,s_i)\in T'} q^{ji}
	=\prod _{(v_i,v_j)\in T} \frac{p^{ij}}{w_{in}(v_j)} 
	\]
	Since in the out-tree $T$, the root node $v_r$ does not have an in-edge, and each of the other nodes in the out-tree has exactly one in-edge, the denominator is a multiplication of all the $w_{in}(v_i)$ values except for $w_{in}(v_r)$:
	\[ 
	w(T') =\frac{1}{\prod _{v_i\in A' , i\neq r} w_{in}(v_i)} \cdot \prod _{(v_i,v_j)\in T} p^{ij}
	\]
	\[
	=\frac{1}{\prod _{v_i\in A' , i\neq r} w_{in}(v_i)} \cdot w(T).
	\]
	
	Dividing both sides by $w_{in}(v_r)$ gives:
	\begin{equation}
		\frac{w(T')}{w_{in}(v_r)} = \frac{w(T)}{ \prod_{v_i\in A'} w_{in}(v_i)} = \alpha\cdot w(T),
	\end{equation}
	where $\alpha=\frac{1}{\prod_{v_i\in A'} w_{in}(v_i)}$, a value that does not depend on $v_r$. 
	
	Now, for $\Pi$ that is calculated in stage $(2)$ of the no-loops method, we get that for every $1\leq i\leq |A'|$, 
	$
	\Pi_i = \sum_{T'\in IT_{i,S_M}} w(T'),
	$
	according to Theorem \ref{TreeThm}.
	Therefore,
	\[
	\Pi^{corr}_i= \frac{\Pi_i}{w_{in}(v_i)} =  \sum_{T'\in IT_{i,S_M}} \frac {w(T')}{w_{in}(v_i)} 
	\]
	\[
	=  \sum_{T\in OT_{i,A'}} \alpha\cdot w(T) =\alpha \cdot \Gamma_i.
	\]
	

\end{proof}
It can be concluded that $\hat{\Pi}^{corr} = \hat{\Gamma}$.
Therefore, by using the no-loops method for the conversion and selecting the node $v_i$ with the maximal $\hat{\Gamma}_i$, we obtain a good estimation for $v^*$.

\begin{figure}
	\centering
	\includegraphics[width= 8cm]{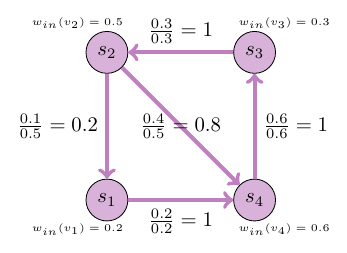}
	\caption{The Markov chain that is obtained by stage (1) of the no-loops method for the social network graph in \ref{fig:complex_ex}.}
	\label{fig no-loops mc}
\end{figure}

To demonstrate the no-loops method and Theorem \ref{NoLoop} we return to the example in Figure \ref{fig:complex_ex}.
The Markov chain that is obtained by stage (1) of the no-loops method for the social network graph in \ref{fig:complex_ex}, is shown in Figure \ref{fig no-loops mc}.
The stationary distribution of this Markov chain is $\Pi=(0.0625,0.3125,0.3125,0.3125)$.
Therefore,
$\Pi^{corr}=(\frac{0.0625}{0.2},\frac{0.3125}{0.5},\frac{0.3125}{0.3},\frac{0.3125}{0.6}).$ 
Finally, after normalization we obtain
$\hat{\Pi}^{corr}=(0.125,$ $0.25,$ $0.417,$ $0.208)$,
which is equal to the output we obtained with the self-loops method.

%






\section{Experiments}

For the evaluation of the performance of the self-loops and the no-loops methods, and comparing them to other baselines heuristics, we use $14$ types of directed random graphs that have diffusion probabilities on their edges\footnote{The code is available at: \url{https://github.com/noamhazon/source-detection}
}, as well as $9$ real-world directed networks from the Social category of the Konect database\footnote{\url{http://konect.cc/networks/}}.
Each of the random graphs is composed using the tuple $(n, Density, p_{range})$: $n$ is the number of nodes, $Density$ is the probability that a directed edge exists between any two nodes $(v_i,v_j)$, and for every directed edge $(v_i,v_j)$, $p^{ij}$ is a random number uniformly drawn from the range $[0,p_{range}]$. Table \ref{tab:graphs} summarizes the parameters of the $14$ random graph types.

\begin{table} [hbpt] 
	\centering
	\caption{Graphs types.}
	\label{tab:graphs}
	\begin{tabular}{l c c c r}\toprule
		Graphs & n & Density &$p_{range}$ & Average no. of edges \\ \midrule
		$G_1$& 500	& 0.1 & 0.0416 & 24957\\
		$G_2$& 1000 & 0.1 & 0.0204 & 99,909\\
		$G_3$& 2000	& 0.1 & 0.0101 & 399,785\\
		$G_4$& 3000	& 0.1 & 0.0071 & 899,721\\
		$G_5$& 4000 & 0.1 & 0.0052 & 1,599,576\\
		$G_6$& 5000 & 0.1 & 0.0041 & 2,499,461\\
		\midrule
		$G_7$&  500  &	0.0416 &	0.1 &	10,404\\
		$G_8$&  1000 &	0.02 &	    0.1	&   20,022\\
		$G_9$&  2000 &	0.0101 &	0.1	&    40,394\\
		$G_{10}$&3000 &	0.0067 &	0.1	&   60,392\\
		$G_{11}$&4000 &	0.0052 &	0.1	&   84,185\\
		$G_{12}$&5000 &	0.0041 &	0.1	&   104,138\\
		$G_{13}$&10000 &	0.002 &	0.1	&   204,092\\
		$G_{14}$&15000 &	0.0013 & 0.1 &	304,001\\
		\bottomrule
	\end{tabular}
\end{table}

Note that $Density$ and $p_{range}$ are chosen such that the average weighted out-degree of each node is slightly greater than $1$. This encourages the diffusion not to be too small on the one hand, but not too large (i.e., including almost the entire graph) on the other.
Similarly, for the real-world networks, we ensured that the average weighted out-degree is slightly greater than $1$.

%
%

For each type of random graph we sampled $1000$ graphs, and for each of these sampled graphs we simulated a single diffusion according to the IC model from a random source node. If the diffusion resulted in less than $20$ active nodes, or if $A'$ was a singleton, another graph was sampled.
Table \ref{tab:trivial_cases_random} summarizes the number of graphs sampled for each type of random graph.

For the real-world networks, we simulated $1000$ diffusions, each starting from a randomly selected source node. If the diffusion resulted in less than $20$ active nodes, or if $A'$ was a singleton, another diffusion was simulated.
Table \ref{tab:trivial_cases_real} summarizes the number of diffusions simulated on each real-world network. Since in the YouTube friends network all diffusions resulted in either too small a diffusion or $|A'|=1$, this network was removed from all further analysis.

\begin{table}[hbpt]
	\centering
	\caption{The number of graphs sampled for each type of random graph. Note that since we require 1000 graph samples with a diffusion with at least 20 active nodes and $|A'|>1$, the total number of samples equals the number of diffusions with less than 20 active nodes plus the number of diffusions with $|A'|>1$ plus 1000.}
	\label{tab:trivial_cases_random}
	\begin{tabular}{l|c | c |c}
		\toprule
		\textbf{Graph type} & \textbf{Total no. }&\textbf{Less than 20 }& \textbf{Diffusions}\\
		
		& \textbf{ of samples}&\textbf{active nodes}& \textbf{with $|A'|=1$} \\
		\midrule
		$G_1$         & 5,232  & 4,199   & 33  \\
		$G_2$         & 5,242  & 4,223   & 19  \\
		$G_3$         & 5,392   & 4,370   & 22  \\
		$G_4$         & 4,432   & 3,420   & 12  \\
		$G_5$         & 4,579   &  3,562  & 17  \\
		$G_6$         & 5,133   &  4,114  & 19  \\
		$G_7$         & 5,223    &  4,092   & 131 \\
		$G_8$         & 7,635   &  6,201 & 434  \\
		$G_9$         & 10,681  &   8,755 & 926  \\
		$G_{10}$         & 12,631  &  10,415 & 1,216  \\
		$G_{11}$         & 8,401  & 6,582  &819  \\
		$G_{12}$         & 11,346  &  9,022 & 1,324 \\
		$G_{13}$         & 22,348 &  18,306 &3,042  \\
		$G_{14}$         & 41,356 &  34,575 & 5,781 \\
		\bottomrule
	\end{tabular}
\end{table}

\begin{table}[hbpt]
	\centering
	\caption{The number of diffusions simulated on each real-world network. Note that in the YouTube friends network, due to its structure, all diffusions resulted in either too small a diffusion or $|A'|=1$.}
	\label{tab:trivial_cases_real}
	\begin{tabular}{l|c | c |c}
		\toprule
		\textbf{Network name} & \textbf{Total no. }&\textbf{Less than 20 }& \textbf{Diffusions}\\
		
		& \textbf{ of diffusions}&\textbf{active nodes}& \textbf{with $|A'|=1$} \\
		\midrule
		Advogato         & 6,894  & 5,211  & 683 \\
		Digg            & 55,160 & 47,092 & 7068\\
		Epinion trust   & 16,362 & 14,502 & 860 \\
		Facebook friends& 8,803  & 7,566  & 237 \\
		Google plus     & 440,477& 162,966& 276,511\\
		Slashdot        & 23,374 & 19,446 & 292 \\
		Twitter         & 50,174 & 46,329 & 3,671 \\
		Youtube links   & 32,473 & 30,396 & 1,077 \\
		Youtube friends & 100,000  & 20,488 & 79,512\\
		\bottomrule
	\end{tabular}
\end{table}

\begin{table*}[hbpt] 
	\centering
	\small\setlength\tabcolsep{4.5pt}
	\caption{The number of times in which each method finds the correct source node in the \emph{random graphs}. The values are out of 1000 cases in which the number of active nodes is at least 20 and $|A'|>1$.}
	\label{tab:results_random_graphs}
	\begin{tabular}{l | c c c c c c | c c c c c c c c | l}\toprule
		&	\textbf{$G_1$}	&	\textbf{$G_2$}	&	\textbf{$G_3$}	&	\textbf{$G_4$}	&	\textbf{$G_5$}	&	\textbf{$G_6$}	&	\textbf{$G_7$}	&	\textbf{$G_8$}	&	\textbf{$G_9$}	&	\textbf{$G_{10}$}	&	\textbf{$G_{11}$}	&	\textbf{$G_{12}$}	&	\textbf{$G_{13}$}	&	\textbf{$G_{14}$}	&	\textbf{Average}	\\
		\midrule
		
		No-loops (direct calc.)	&	109	&	101	&	90	&	64	&	78	&	79	&	131	&	150	&	156	&	174	&	95	&	114	&	136	&	160	&	\textbf{116.92}	\\
		
		Naive                    	&	50	&	36	&	52	&	36	&	30	&	39	&	43	&	47	&	57	&	61	&	22	&	39	&	38	&	61	&	43.64	\\
		Random                                        	&	22	&	23	&	14	&	15	&	13	&	13	&	22	&	32	&	32	&	54	&	22	&	34	&	31	&	39	&	26.14\\
		Max out-deg                            	&	33	&	47	&	35	&	17	&	28	&	26	&	47	&	51	&	51	&	61	&	27	&	40	&	32	&	49	&	38.85	\\
		Min in-deg                             	&	69	&	57	&	46	&	33	&	40	&	51	&	44	&	40	&	40	&	58	&	20	&	36	&	33	&	38	&	43.21	\\
		Max (out/in)-deg                  	&	78	&	69	&	52	&	36	&	44	&	62	&	75	&	76	&	61	&	69	&	33	&	41	&	38	&	50	&	56	\\
		IM based                        	&	31	&	47	&	38	&	35	&	18	&	25	&	49	&	52	&	48	&	67	&	20	&	29	&	40	&	53	&	39.42	\\
		Max weight arbo. \cite{amoruso2020contrasting}                  	&	79	&	83	&	73	&	49	&	65	&	63	&	115	&	127	&	133	&	160	&	88	&	101	&	127	&	152	&	101.07	\\
		MCMC \cite{zhai2015cascade}  &	31	&	40   &	31 &	27	&	21 &	33 & 76 &	57	&	59	&	62	&	37	&	46	&	47	&	83	&	46.29	\\
		
		\bottomrule
	\end{tabular}
\end{table*}

\begin{table*}[hbpt] 
	\centering
	\small\setlength\tabcolsep{4.5pt}
	\caption{The number of times in which each method finds the correct source node in the \emph{real-world networks}. The values are out of 1000 cases in which the number of active nodes is at least 20 and $|A'|>1$.}
	\label{tab:results_real_graphs}
	\begin{tabular}{l | c c c c c c c c | l}\toprule
		& Advogato	&	Digg	&	Epinion	&	 Facebook 	&	Google 	&	Slashdot	&	Twitter	&	Youtube 
		&	\textbf{Average}	\\
		& 	&		&	  trust	&	  friendships	&	plus 	&		&		&	links 
		&		\\
		\midrule
		No-loops (direct calc.)	&	222	&	451	&	428	&	354	&	125	&	471	&	241	&	486	&	\textbf{347.25}	\\
		Naive                    	&	139	&	172	&	146	&	176	&	78	&	174	&	149	&	131	&	145.625	\\
		Random                                        	&	52	&	130	&	98	&	89	&	71	&	137	&	115	&	79	&	96.375	\\
		Max out-deg                            	&	39	&	115	&	82	&	76	&	79	&	130	&	132	&	47	&	87.5	\\
		Min in-deg                             	&	70	&	162	&	125	&	117	&	72	&	155	&	115	&	128	&	118	\\
		Max (out/in)-deg                  	&	63	&	132	&	115	&	95	&	86	&	141	&	161	&	154	&	118.375	\\
		IM based                        	&	94	&	309	&	230	&	196	&	120	&	302	&	218	&	273	&	217.75	\\
		Max weight arbo.\cite{amoruso2020contrasting}                  	&	136	&	353	&	329	&	278	&	125	&	380	&	184	&	358	&	267.875	\\
		MCMC \cite{zhai2015cascade}   & 98 &	238	&	221	&	202	&	116	&	280	&	185	&	142	&	185.25	\\
		\bottomrule
	\end{tabular}
\end{table*}

\begin{figure}[hbpt] 
	\centering
	\includegraphics[width=0.8\columnwidth]{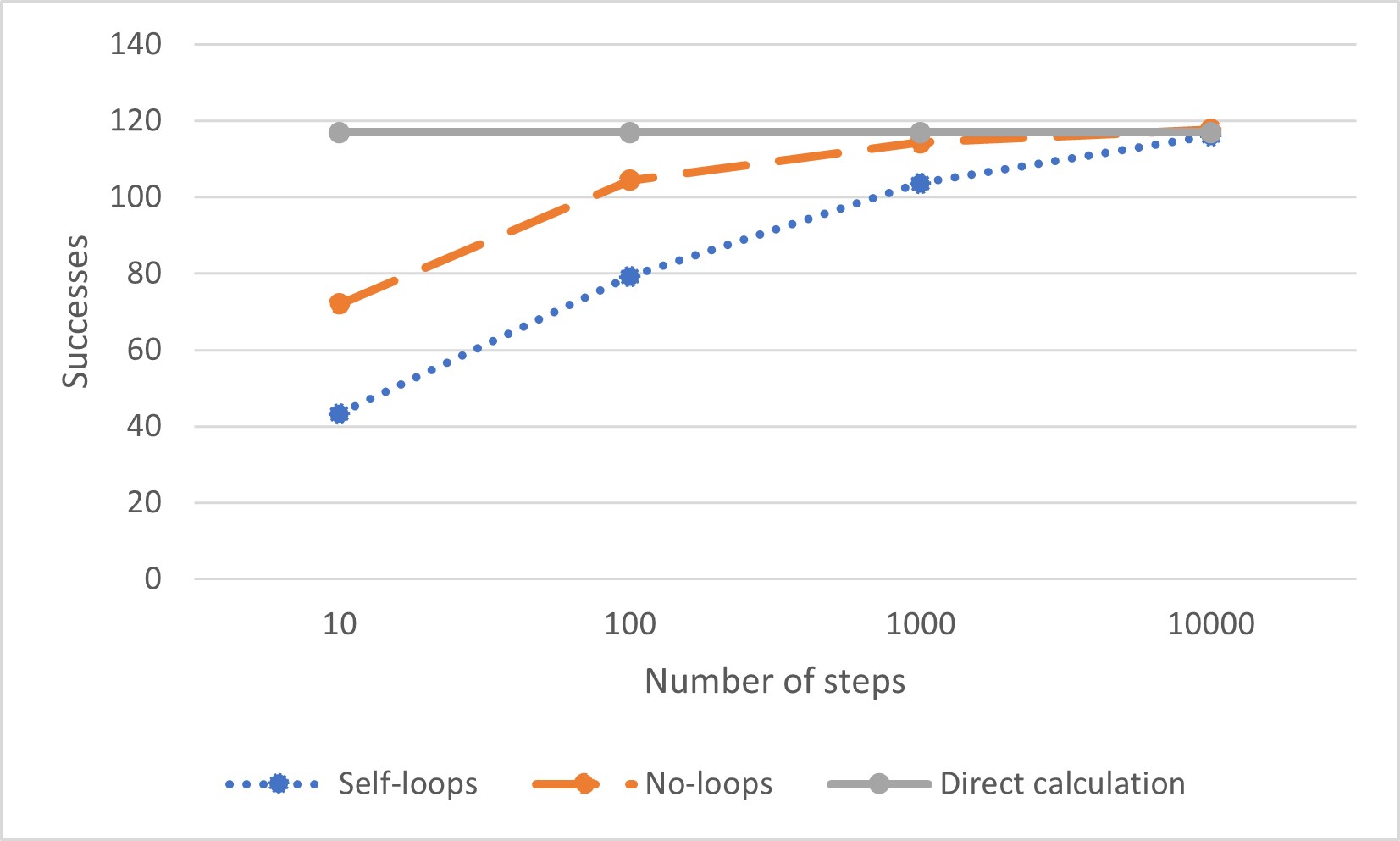}
	\caption{The average number of times in which the no-loops and the self-loops methods, using direct calculation and using random walks 
		with various number of steps, 
		find the correct source node on the \emph{random graphs}. }
	\label{fig:random_walk_chart_random}
\end{figure}

\begin{figure}[hbpt] 
	\centering
	\includegraphics[width=0.8\columnwidth]{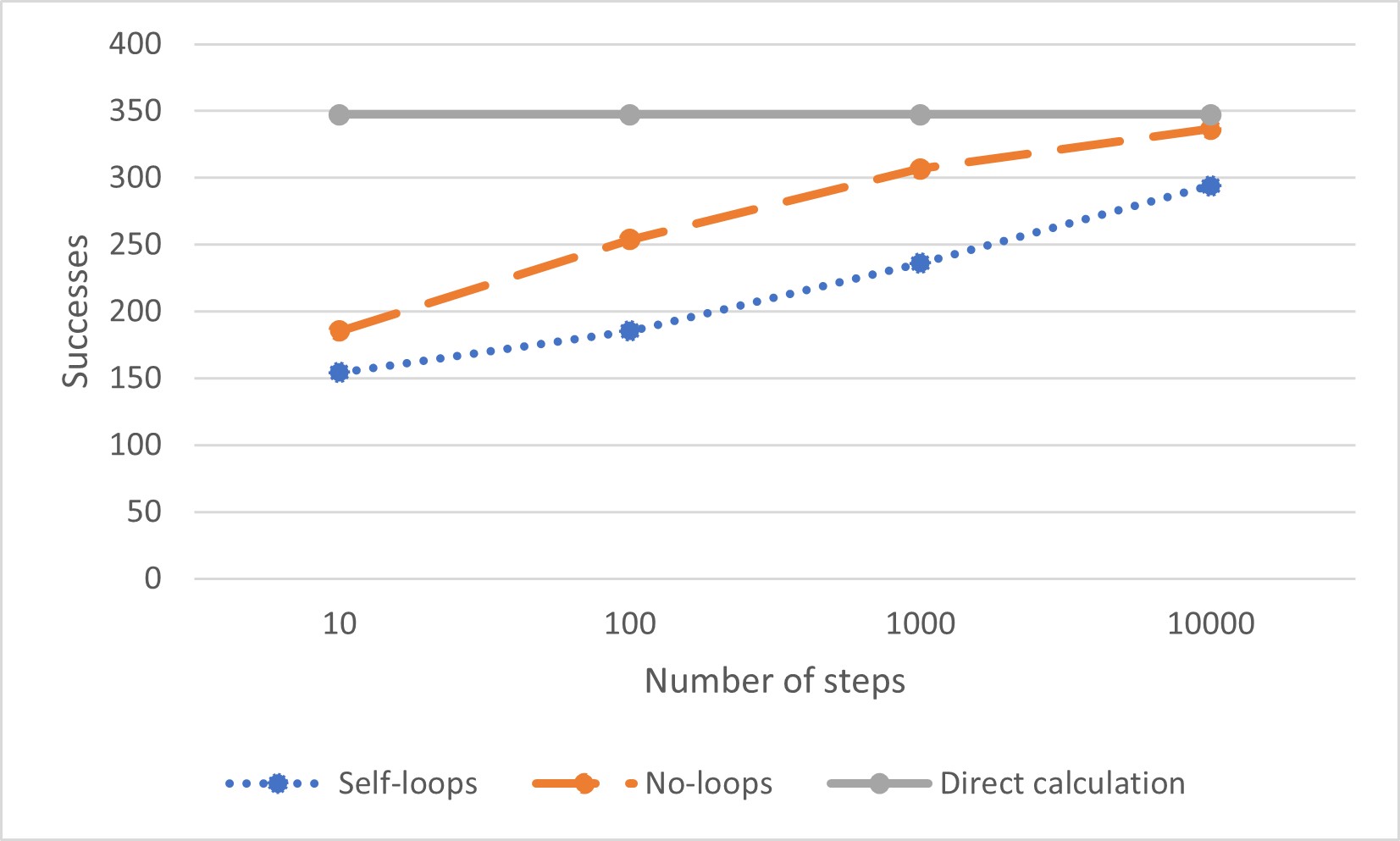}
	\caption{The average number of times in which the no-loops and the self-loops methods, using direct calculation and using random walks, find the correct source node on the \emph{real-world networks}.}
	\label{fig:random_walk_chart_real}
\end{figure}

We begin by evaluating the performance of the self-loops and the no-loops methods with a direct calculation of the stationary distribution, and when the stationary distribution is estimated by random walks with $10$, $100$, $1000$, or $10,000$ steps.
The summary of the results is presented in Figures \ref{fig:random_walk_chart_random} and \ref{fig:random_walk_chart_real} (the detailed results are presented in Tables \ref{tab:results_random_graphs_rand_walk} and \ref{tab:results_real_graphs_rand_walk} in the Appendix). 
As expected, with direct calculation, both methods (self-loops and no-loops) result in the exact same solution, and take a similar time to arrive at this solution. However, with random walks, the no-loops method requires fewer steps than the self-loops method for providing the same performance. 
This is in line with our claim in Section \ref{sec:no-loop} that the self-loops method might require longer walks and will thus be less efficient when the stationary distribution is estimated by sampling. Therefore, one can state that, overall, the no-loops method is superior to the self-loops method.

Next, we evaluate the performance of the no-loops method, using direct calculation, against the following baseline methods:
\begin{itemize}
	\item \textbf{Naive:} The Markov chain approach with the naive conversion method.
	\item \textbf{Random:} A random selection of a node in $A'$. 
	\item \textbf{Max out-degree:} The node with the maximal weighted out-degree is selected. 
	\item \textbf{Min in-degree:} The node with the minimal weighted in-degree is selected.
	\item \textbf{Max (out/in) degree:} The node with the maximal weighted out-degree divided by its weighted in-degree is selected.
	\item \textbf{IM based:} For each node, we simulate $1000$ diffusions, and the node with the maximal average size of the active set is selected.
	\item \textbf{Maximum arborescence \cite{amoruso2020contrasting}:} The node that is the root of the maximum weight spanning out-tree (arborescence) is selected.
	\item \textbf{Markov Chain Monte Carlo (MCMC) \cite{zhai2015cascade}:} 
	The algorithm uses a Markov-chain random walk to sample networks consistent with the observed active set. It then scores each node by how often it is a feasible source across the samples and returns the highest-scoring node.
\end{itemize}


Our results are presented in Tables \ref{tab:results_random_graphs} and \ref{tab:results_real_graphs}.
As can be observed, the no-loops method outperformed the other baseline methods, both on random graphs and real-world networks.
Interestingly, the average $|A'|$ for the random graphs is $196.2$, while the average $|A'|$ for the real-world networks was lower, $126.2$.
Therefore, the average number of times the methods found the correct source (successes) in the random graphs is much lower than in the real-world networks. In addition, there is a substantial difference in the number of correct source-node identifications across the real-world networks. This difference can be attributed to the inherent variations in their structures. For example, the edge density of YouTube links is $4.34$, while the edge density of the Google+ network is only $1.66$.

\section{Conclusions and Future Work}
In this paper, we study the problem of identifying the source of a given diffusion on a network. We use the common IC model, and utilize the MLE principle. Our approach is based on computing the stationary distribution of a Markov chain. Interestingly, with this approach, rather than computing the likelihood of every node to be the source separately, the values for all nodes are derived from the same stationary distribution. We propose two approaches for converting the network to a Markov chain, and demonstrate the effectiveness of one of them, the no-loops method, even when using random walks to estimate the stationary distribution. 

For future work, we would like to extend our Markov chain approach to settings in which not all weights, edges, or even nodes are known (see, for example, \cite{gomez2012inferring,altarelli2014patient,braunstein2019network,wang2025learning}). We note that if there are missing edges or nodes, one must consider all active nodes as possible sources rather than only $A'$.
In addition, we would like to extend our approach to other diffusion models, such as the linear threshold model \cite{kempe2003maximizing}, and the continuous time independent cascade model \cite{kim2014ct}.

\section*{Acknowledgement}	This research has been partly supported by the Israel Science Foundation under grant 1092/24, and by the Ministry of Science and Technology, Israel.

\bibliographystyle{ACM-Reference-Format} 
\bibliography{sample}


\onecolumn
\appendix
\section*{Appendix}
\label{sec:appx}
\begin{table}[hbpt] 
	\centering
	\small\setlength\tabcolsep{4.5pt}
	\caption{The number of times in which each method finds the correct source node in the \emph{random graphs}. The values are out of 1000 cases in which the number of active nodes is at least 20 and $|A'|>1$.}
	\label{tab:results_random_graphs_rand_walk}
	\begin{tabular}{l | c c c c c c | c c c c c c c c | l}\toprule
		&	\textbf{$G_1$}	&	\textbf{$G_2$}	&	\textbf{$G_3$}	&	\textbf{$G_4$}	&	\textbf{$G_5$}	&	\textbf{$G_6$}	&	\textbf{$G_7$}	&	\textbf{$G_8$}	&	\textbf{$G_9$}	&	\textbf{$G_{10}$}	&	\textbf{$G_{11}$}	&	\textbf{$G_{12}$}	&	\textbf{$G_{13}$}	&	\textbf{$G_{14}$}	&	\textbf{Average}	\\
		\midrule
		Self-loops (direct calc.)	&	109	&	101	&	90	&	64	&	78	&	79	&	131	&	150	&	156	&	174	&	95	&	114	&	136	&	160	&	\textbf{116.92}	\\
		10 steps	&	46	&	36	&	39	&	22	&	23	&	35	&	44	&	48	&	59	&	63	&	25	&	47	&	59	&	60	&	43.28	\\
		100 steps	&	69	&	80	&	63	&	46	&	51	&	45	&	92	&	111	&	99	&	108	&	54	&	72	&	99	&	120	&	79.21	\\
		1000 steps	&	97	&	95	&	82	&	45	&	67	&	71	&	124	&	135	&	153	&	154	&	71	&	95	&	118	&	144	&	103.64	\\
		10000 steps	&	110	&	103	&	90	&	60	&	76	&	72	&	126	&	154	&	155	&	176	&	86	&	114	&	132	&	171	&	116.07	\\
		\midrule
		
		No-loops (direct calc.)	&	109	&	101	&	90	&	64	&	78	&	79	&	131	&	150	&	156	&	174	&	95	&	114	&	136	&	160	&	\textbf{116.92}	\\
		10 steps	&	59	&	66	&	66	&	40	&	41	&	48	&	81	&	97	&	93	&	96	&	52	&	69	&	90	&	110	&	72	\\
		100 steps	&	96	&	89	&	89	&	61	&	67	&	73	&	116	&	133	&	138	&	155	&	80	&	96	&	123	&	146	&	104.428	\\
		1000 steps	&	102	&	106	&	81	&	62	&	79	&	69	&	129	&	151	&	154	&	171	&	90	&	113	&	135	&	159	&	114.35	\\
		10000 steps	&	109	&	104	&	90	&	63	&	82	&	77	&	136	&	149	&	158	&	177	&	95	&	114	&	134	&	160	&	117.71	\\
		
		\bottomrule
	\end{tabular}
\end{table}

\begin{table}[hbpt] 
	\centering
	\small\setlength\tabcolsep{4.5pt}
	\caption{The number of times in which each method finds the correct source node in the \emph{real-world networks}. The values are out of 1000 cases in which the number of active nodes is at least 20 and $|A'|>1$.}
	\label{tab:results_real_graphs_rand_walk}
	\begin{tabular}{l | c c c c c c c c | l}\toprule
		& Advogato	&	Digg	&	Epinion	&	 Facebook 	&	Google 	&	Slashdot	&	Twitter	&	Youtube 
		&	\textbf{Average}	\\
		& 	&		&	  trust	&	  friendships	&	plus 	&		&		&	links 
		&		\\
		\midrule
		Self-loops (direct calc.)	&	222	&	451	&	428	&	354	&	125	&	471	&	241	&	486	&	\textbf{347.25}	\\
		10 steps	&	68	&	216	&	161	&	146	&	89	&	246	&	167	&	141	&	154.25	\\
		100 steps	&	86	&	237	&	205	&	170	&	98	&	289	&	230	&	168	&	185.375	\\
		1000 steps	&	111	&	300	&	270	&	233	&	116	&	358	&	230	&	270	&	236	\\
		10000 steps	&	175	&	383	&	363	&	308	&	118	&	418	&	247	&	341	&	294.125	\\
		\midrule
		No-loops (direct calc.)	&	222	&	451	&	428	&	354	&	125	&	471	&	241	&	486	&	\textbf{347.25}	\\
		10 steps	&	88	&	248	&	216	&	179	&	82	&	301	&	167	&	203	&	185.5	\\
		100 steps	&	132	&	311	&	286	&	252	&	121	&	389	&	253	&	287	&	253.875	\\
		1000 steps	&	191	&	398	&	384	&	307	&	128	&	444	&	236	&	367	&	306.875	\\
		10000 steps	&	214	&	441	&	423	&	344	&	126	&	468	&	236	&	441	&	336.625	\\
		
		\bottomrule
	\end{tabular}
\end{table}

\end{document}